\documentclass[english,aps,preprint]{revtex4}
\usepackage[T1]{fontenc}
\usepackage[utf8]{inputenc}
\usepackage{amsmath}
\usepackage{amssymb}
\usepackage{circuitikz}
\usepackage{xcolor}
\usepackage{graphicx}
\usepackage{subcaption}
\makeatletter
\@ifundefined{textcolor}{}
{%
 \definecolor{BLACK}{gray}{0}
 \definecolor{WHITE}{gray}{1}
 \definecolor{RED}{rgb}{1,0,0}
 \definecolor{GREEN}{rgb}{0,1,0}
 \definecolor{BLUE}{rgb}{0,0,1}
 \definecolor{CYAN}{cmyk}{1,0,0,0}
 \definecolor{MAGENTA}{cmyk}{0,1,0,0}
 \definecolor{YELLOW}{cmyk}{0,0,1,0}
}

\usepackage{babel}

\makeatother

\usepackage{babel}
\begin{document}
\global\long\def\abs#1{\left|#1\right|}%
\global\long\def\ket#1{\left|#1\right\rangle }%
\global\long\def\bra#1{\left\langle #1\right|}%
\global\long\def\half{\frac{1}{2}}%
\global\long\def\partder#1#2{\frac{\partial#1}{\partial#2}}%
\global\long\def\comm#1#2{\left[#1,#2\right]}%
\global\long\def\vp{\vec{p}}%
\global\long\def\vpp{\vec{p}'}%
\global\long\def\dt#1{\delta^{(3)}(#1)}%
\global\long\def\Tr#1{\textrm{Tr}\left\{  #1\right\}  }%
\global\long\def\Real#1{\mathrm{Re}\left\{  #1 \right\}  }%
\global\long\def\braket#1{\langle#1\rangle}%
\global\long\def\escp#1#2{\left\langle #1|#2\right\rangle }%
\global\long\def\elmma#1#2#3{\langle#1\mid#2\mid#3\rangle}%
\global\long\def\ketbra#1#2{|#1\rangle\langle#2|}%

\preprint{APS/123-QED}
\title{Quantum simulation of classical RCL circuits via complexification and Schrödingerization}
\author{Huner Fanchiotti}
\affiliation{IFLP/CONICET and Departamento de Física, Universidad Nacional de La
Plata (UNLP), C.C.67, 1900, La Plata, Argentina }
\author{Carlos A. García Canal}
\affiliation{IFLP/CONICET and Departamento de Física, Universidad Nacional de La
Plata (UNLP), C.C.67, 1900, La Plata, Argentina }
\author{Miguel A. Mayosky}
\affiliation{Instituto de Investigaciones en Electrónica, Control y Procesamiento
de Señales - LEICI (UNLP-CONICET), Facultad de Ingeniería, Universidad
Nacional de La Plata, Argentina}
\author{Alejandro L. Veiga}
\affiliation{Instituto de Investigaciones en Electrónica, Control y Procesamiento
de Señales - LEICI (UNLP-CONICET), Facultad de Ingeniería, Universidad
Nacional de La Plata, Argentina}
\author{Armando Pérez}
\affiliation{Departamento de Física Teórica-IFIC. Universidad de Valencia-CSIC.E-46100,
Burjassot (Valencia), Spain}
\author{Rafael Gómez-Lurbe}
\affiliation{Departamento de Física Teórica-IFIC. Universidad de Valencia-CSIC.E-46100,
Burjassot (Valencia), Spain}
\author{Vicente Vento}
\affiliation{Departamento de Física Teórica-IFIC. Universidad de Valencia-CSIC.E-46100,
Burjassot (Valencia), Spain}
\begin{abstract}
The decomplexification procedure establishes a mathematical equivalence
between finite dimensional quantum dynamics and classical dynamics
by mapping the complex quantum domain onto the real classical one.
Its inverse process, complexification --- which lifts a classical
dynamics to an equivalent quantum formulation --- is of great interest,
particularly because it enables the quantum simulation of equations
that are of classical relevance. In this way, one can prepare the
solution of classical differential equations to be implemented using
quantum computation tools. In the present work, we recall the decomplexification
procedure and also discuss several distinct strategies for performing
the complexification of classical equations, so as to obtain the corresponding
quantum-equivalent dynamics.
\end{abstract}
\maketitle

\section{Introduction}

The simulation of dynamical processes is central to science and, in
particular, to physics as it allows for a better understanding of
the laws governing those processes. The development of quantum computers
promises a revolution in this field. On the one hand, quantum computers,
as anticipated by Yuri Manin \cite{Manin1980} and shortly after by
Richard Feynman \cite{Feynman1982} offer the possibility of surpassing
the inherent difficulty to simulate quantum systems as the size of
the system grows \cite{Cirac2012,https://doi.org/10.1002/wcms.70020,miessen2025digitalquantumsimulationmanybody}.
On the other hand, quantum computers are also candidates to boost
up, in an unprecedented way, calculations that are nowadays performed
by classical computers \cite{PhysRevResearch.2.043102,Tennie2025}.

A crucial ingredient behind both topics is understanding the relationship
between quantum and classical dynamics. The proof of the mathematical
equivalence between quantum and classical dynamics of systems of finite
dimension \cite{doi:10.1098/rspa.2016.0615} is based upon the decomplexification
procedure \cite{Arnold2012,arnold1992ode}. By means of this technique,
one solves the difficulty in comparing both dynamics due to the fact
that Quantum Mechanics is defined in a complex space, while Classical
Mechanics lives in a real space. The decomplexification procedure
of a complex vector space is valid because any complex vector space
can be rewritten as a real vector space \cite{bookmath}. In doing
this, one has to omit the multiplication of a vector by a complex
number because only the product by real numbers is allowed.

The decomplexification procedure, as well as its inverse one, complexification,
has a precise mathematical meaning. Both procedures are briefly presented
in the next section and related to the problem of equivalence between their corresponding dynamics.

The equivalence just described has two main consequences. On the one
hand it provides the possibility of establishing a connection between
quantum systems and classical oscillators, in particular with electric
resonant circuits; on the other hand, complexification allows quantum
simulations of classical equations. Therefore, this analysis leads
to the possibility of simulating in the laboratory (or numerically)
the time evolution of equivalent quantum systems \cite{Fanchiotti2022,Fanchiotti2023}.
Moreover, inversely one may apply quantum computing technology to
simulate classical systems.

%In producing the analysis of several examples of complexification of classical dynamics, we %discovered that the real and
%imaginary parts of the quantum wave function one obtains when considering
%the mentioned mathematical equivalence, are not independent. In fact they result Hilbert
%transform conjugate. This fact triggered a new critical analysis of the physical
%content of the mathematical equivalence between quantum and classical
%dynamics. This is treated in detail at the end of this presentation.

The next section is devoted to a presentation of the complexification
and decomplexification procedures. There, two cases are treated in
detail, to fix ideas and to show the potentiality of both procedures.
In section \ref{SCH} we include, for completeness and for the scope
of comparison, a recently proposed method \cite{PhysRevLett.133.230602}
for "Schrödingerisation" of classical differential equations, which
certainly implies the complexification of the original ones. This
method appears very useful for computing solutions of the classical
equations in the quantum domain. We end by drawing some conclusions
of our analysis.

\section{ Decomplexification and Complexification Procedures}

We next present a brief mathematical description of the 
decomplexification and complexification processes used in the analysis of differential
equations to treat those cases where a real equation has complex eigenvalues.
These cases are of particular interest because the real space is not
enough to account for the equation and its solutions.

We first summarize the decomplexification formalism, which
defines a classical analog of a quantum system, to establish the connection
between quantum and classical dynamics. This can be ultimately presented
as an analog description of quantum systems in terms of oscillatory
electric circuits.

The decomplexification procedure \cite{Arnold2012} of quantum Hamiltonians
is presented here for the simple case of a $2\times2$ Hamiltonian
just to fix ideas, but this procedure works for any Hamiltonian of
finite dimension. The starting Schrödinger equation is

\begin{equation}
i\,\frac{\partial}{\partial t}\Psi=H\,\Psi.
\end{equation}
where we use $\hbar=1$. The Hamiltonian $H$ is a hermitian $2\times2$
matrix, and $\Psi$ is a two component vector.

Using the notation 
\[
Re(\Psi_{i})=x_{i}\,\,\,;\,\,\,Im(\Psi_{i})=y_{i}\,\,;\,i=1,2\,,
\]
\begin{equation}
Re(H_{ij})=f_{ij}\,\,\,;\,\,\,Im(H_{ij})=g_{ij}\,\,;\,i,j=1,2\,,
\end{equation}
after separating real and imaginary parts and using matrix notation
we get for the Schrödinger equation

\begin{equation}
\left[\begin{array}{c}
\dot{x}_{1}\\
\dot{y}_{1}\\
\dot{x}_{2}\\
\dot{y}_{2}
\end{array}\right]=\left[\begin{array}{cccc}
\;\;\,g_{11} & f_{11} & \;\;\,g_{12} & f_{12}\\
-f_{11} & g_{11} & -f_{12} & g_{12}\\
\;\;\,g_{21} & f_{21} & \;\;\,g_{22} & f_{22}\\
-f_{21} & g_{21} & -f_{22} & g_{22}
\end{array}\right]\left[\begin{array}{c}
x_{1}\\
y_{1}\\
x_{2}\\
y_{2}
\end{array}\right].\label{ydot}
\end{equation}
Defining 
\begin{equation}
A=\left[\begin{array}{cc}
g_{11} & g_{12}\\
g_{21} & g_{22}
\end{array}\right]\,\,\,\,\,\,;\,\,\,\,\,\,B=\left[\begin{array}{cc}
f_{11} & f_{12}\\
f_{21} & f_{22}
\end{array}\right]\label{AyB}
\end{equation}
and 
\begin{equation}
x=\left[\begin{array}{c}
x_{1}\\
x_{2}
\end{array}\right]\,\,\,\,\,\,;\,\,\,\,\,\,y=\left[\begin{array}{c}
y_{1}\\
y_{2}
\end{array}\right]\label{xy12}, 
\end{equation}
the system of first order differential equations that has equivalent
mathematical information as the starting quantum one becomes 
\begin{eqnarray}
\dot{x} & = & A\,x+B\,y\nonumber \\
\dot{y} & = & A\,y-B\,x
\end{eqnarray}

These equations can be transformed into two separated second order
differential equations for $x$ and $y$ that can immediately be identified
with classical oscillators, namely

\begin{eqnarray}
\ddot{x} & = & (A+B\,A\,B^{-1})\,\dot{x}-(B\,A\,B^{-1}\,A\,+B\,B)\,x,\nonumber \\
\label{ex}
\end{eqnarray}
and an entirely similar one for $y$.

In Eq. (7), the matrix A appearing in the coefficient of the first derivative of x does it with the same sign as the second term of the same coefficient. This allows to describe in the formalism the dissipative part of the dynamics generated by an imaginary part of an effective Hamiltonian, thus making it possible to study decaying states. Conversely, the dynamics of a PT-symmetric Hamiltonian could be described if both terms had opposite signs.

As mentioned earlier, the above procedure can be formally extended
to Hamiltonians of finite arbitrary dimension \cite{doi:10.1098/rspa.2016.0615}.

We now provide elements for introducing complexification, the opposite
procedure to decomplexification. Namely, to build up the Hamiltonian
$H$ or the complex differential equation of Schrödinger type, from
the knowledge of what should be the equivalent classical dynamics.

We present first a procedure that starts from the set of differential
equations of second order under consideration (in the example that
follows, the equation is related to electric oscillators). These equations
can be transformed, in the standard way, into a set of first order
differential equations that are written in matrix notation. The next
step is to find the eigenvalues of this matrix characterizing the
system. Notice that this real matrix, the state matrix, can have real
or complex conjugate eigenvalues. Since these eigenvalues can be complex,
from a mathematical point of view, one has to extend the differential
equation from the corresponding real space to a complex space. A way
to do this is to find the matrix that diagonalizes the state matrix.
Using the diagonalizing matrix two complex conjugate equations are
generated, which show how one can go from a $R^{2n}$ space to a $C^{n}$
space. By taking one of them one can construct a Schrödinger type
equation simply by multiplying the equation by $i$. In this way,
one obtains a quantum equivalent Hamiltonian. This proposal, involving
the solution of algebraic equations, although it can be generalized
for larger dimensions, implies the solution of complex algebraic equations
that in most cases have to be treated by numerical methods.

To fix ideas, we present a couple of examples. Let us first consider
a very simple case of the differential equation of only one classical damped
oscillator, namely, a resonant circuit that satisfies 
\begin{equation}
\frac{d^{2}}{dt^{2}}V+2\gamma\frac{d}{dt}V+\omega^{2}V=0\label{ho}
\end{equation}
with some initial conditions, non relevant for the moment. This equation
can be transformed into a system of first order differential equations
as follows. Calling 
\[
x=V\,\,\,\,;\,\,\,\,y=\frac{dV}{dt}
\]
equation ({\ref{ho}}) becomes 
\begin{equation}
\frac{d}{dt}\left[\begin{array}{c}
x\\
y
\end{array}\right]=\left[\begin{array}{cc}
0 & 1\\
-\omega^{2} & -2\gamma
\end{array}\right]\left[\begin{array}{c}
x\\
y
\end{array}\right]\label{eq:oneoscillator}
\end{equation}
which has the form 
\begin{equation}
\frac{d}{dt}\left(\overset{\rightarrow}{x}\right)=A\overset{\rightarrow}{x}
\end{equation}
where, in general, the state vector $\overset{\rightarrow}{x}$ has
$n$ components ($2$ in this example) in a real space $\left(R^{n}\right)$
and $A$ is a $\left(n\times n\right)$ matrix with real elements
($2\times2$ in this case)

For the example under consideration, the state vector is $\left[\begin{array}{c}
x\\
y
\end{array}\right]$ and 
\begin{equation}
A=\left[\begin{array}{cc}
0 & 1\\
-\omega^{2} & -2\gamma
\end{array}\right] .
\end{equation}

The next step is to find the eigenvalues of $A$, which are 
\begin{equation}
\lambda_{1}=-\gamma+i\sqrt{\omega^{2}-\gamma^{2}}\,\,\,\,;\,\,\,\,\lambda_{2}=-\gamma-i\sqrt{\omega^{2}-\gamma^{2}}
\end{equation}
clearly being complex conjugate of each other when $\omega^{2}\geq\gamma^{2}$.
Then, these eigenvalues escape from the real space, and the same happens
with the eigenstates of $A$ 
\begin{equation}
\left[\begin{array}{c}
\lambda_{1}\\
\omega^{2}
\end{array}\right]\,\,\,;
\end{equation}
\begin{equation}
\left[\begin{array}{c}
\lambda_{2}\\
\omega^{2}
\end{array}\right]\,\,\,.
\end{equation}

Consequently, one has to extend the differential equation from the
real space $R^{2}$ to the complex space ($C^{2}$, in this case).
To do this it is enough to find the matrix that diagonlizes $A$, namely
\begin{equation}
S=\frac{1}{2\sqrt{\omega^{2}-\gamma^{2}}}\left[\begin{array}{cc}
-i\omega^{2} & i\lambda_{2}\\
i\omega^{2} & -i\lambda_{1}
\end{array}\right]
\end{equation}

Then, 
\begin{equation}
\frac{d}{dt}S\left[\begin{array}{c}
x\\
y
\end{array}\right]=S\left[\begin{array}{cc}
0 & 1\\
-\omega^{2} & -2\gamma
\end{array}\right]S^{-1}S\left[\begin{array}{c}
x\\
y
\end{array}\right]
\end{equation}
which, after introducing the complex eigenstates, gives rise to 
\begin{equation}
S\left[\begin{array}{c}
x\\
y
\end{array}\right]=\left[\begin{array}{cc}
-i\omega^{2} & i\lambda_{2}\\
i\omega^{2} & -i\lambda_{1}
\end{array}\right]\left[\begin{array}{c}
x\\
y
\end{array}\right]
\end{equation}
\begin{equation}
=\left[\begin{array}{c}
y\left(1+\sqrt{\omega^{2}-\gamma^{2}}\right)-i\left(\gamma y+x\omega^{2}\right)\\
y\left(1+\sqrt{\omega^{2}-\gamma^{2}}\right)+i\left(y\gamma+x\omega^{2}\right)
\end{array}\right]=\left[\begin{array}{c}
z\\
z^{\ast}
\end{array}\right].
\end{equation}
Being complex conjugate of each other, one can proceed taking into account only
one of them and one has complexified the real equation going from
$R^{2}$ to the complex space $C^{1}$. The equation now is 
\begin{equation}
\frac{d}{dt}z=\left(-\gamma\allowbreak+i\sqrt{\omega^{2}-\gamma^{2}}\right)z=\lambda_{1}z,
\end{equation}
which can be read as a Schrödinger equation with the, as expected,
non hermitian Hamiltonian 
\begin{equation}
H=i\lambda_{1}
\end{equation}
a complex number in this trivial case.

Now we present in detail the case of complexification of the differential
equations that correspond to two $PT$-symmetric resonant circuits, 
coupled by gyrator, represented in Fig. 1, namely 
\begin{equation}
\overset{\cdot\cdot}{V}_{1}-\gamma\overset{\cdot}{V}_{1}+\omega^{2}V_{1}+\gamma_{g}\overset{\cdot}{V}_{2}=0
\end{equation}
\begin{equation}
\overset{\cdot\cdot}{V}_{2}+\gamma\overset{\cdot}{V}_{2}+\omega^{2}V_{2}-\gamma_{g}\overset{\cdot}{V}_{1}=0 ,
\end{equation}
where 
\begin{eqnarray}
\begin{array}{ccc}
\omega^{2}_{0}=\frac{1}{LC}\,\,\,\,;\,\,\,\,\gamma=\frac{1}{RC}\,\,\,\,;\,\,\,\,\gamma_{g}=\frac{G_{g}}{C}\end{array} .
\end{eqnarray}

Recall that a gyrator is a non-reciprocal passive element of two ports
with a given conductance \cite{Telle48} 

\begin{figure}[h!]
\centering
    \begin{circuitikz}[scale = 1, transform shape]
      \draw	(0,2.1) to [short] (-1,2.1) to [R=$-R$] (-1,0)to [short] (0,0);	
			\draw	(0.25,2.1) to[L=$L$, ] (0.25,0) ;		
			\draw	(0,2.1) to[short] (2,2.1);
      \draw (1.5,2.1) to [C=$C$, ] (1.5,0) to[short] (0,0) to [short] (2,0);
			\draw (4.5,2.1)	to[C=$C$, ](4.5,0) to[short](6,0) to [short] (4,0);
			\draw (3,1.05) node[gyrator]{} node[above=.5cm]{$G_g$};
			\draw (4,2.1)	to[short] (5.75,2.1) to[L=$L$, ](5.75,0);
			\draw(5.75,2.1) to [short] (7,2.1) to [R=$R$]  (7,0) to [short] (6,0);
			\draw(1,0) node[ground]{};
			\draw(5,0) node[ground]{};
			\draw(1,2.1) node [label=$V_1$]{};
			\draw(5,2.1) node [label=$V_2$]{};
    \end{circuitikz}
\caption{Gyrator-coupled oscillator circuit.}
\label{circuit}
\end{figure}
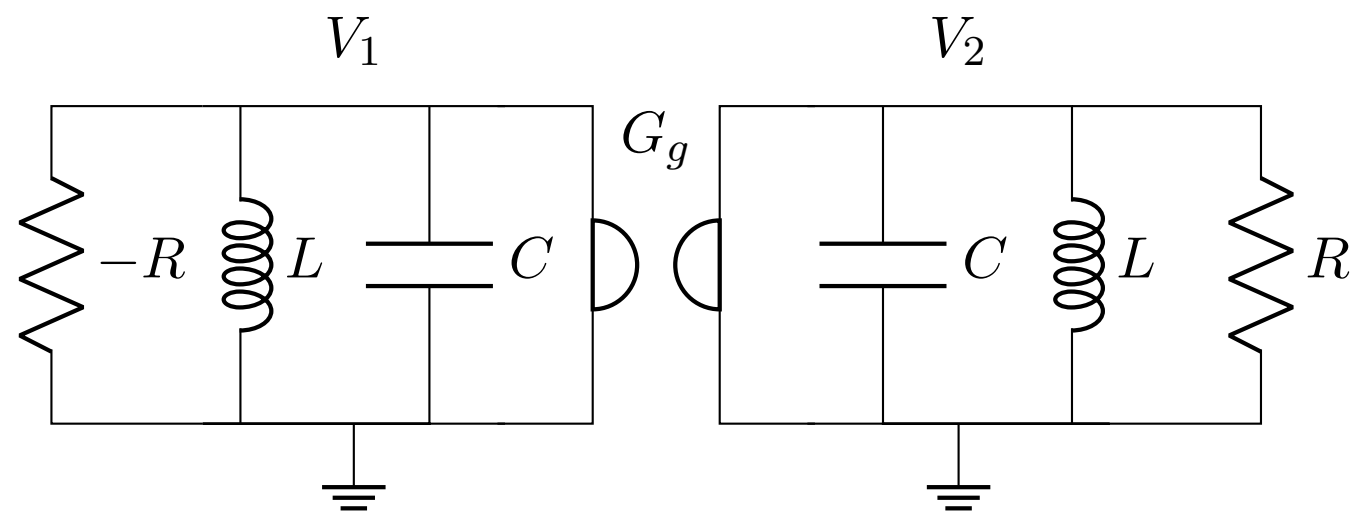

Calling as before $x$ and $y$ to the potentials and their derivatives,
one gets 
\begin{equation}
\frac{d}{dt}\left[\begin{array}{c}
x_{1}\\
x_{2}\\
y_{1}\\
y_{2}
\end{array}\right]=\left[\begin{array}{cccc}
0 & 0 & 1 & 0\\
0 & 0 & 0 & 1\\
-\omega^{2} & 0 & \gamma & \gamma_{g}\\
0 & -\omega^{2} & -\gamma_{g} & -\gamma
\end{array}\right]\left[\begin{array}{c}
x_{1}\\
x_{2}\\
y_{1}\\
y_{2}
\end{array}\right]=A\,\left[\begin{array}{c}
x_{1}\\
x_{2}\\
y_{1}\\
y_{2}
\end{array}\right],
\end{equation}
where the matrix $A$ has the characteristic polynomial 
\begin{equation}
det(A-X\,I)=X^{4}+\left(-\gamma^{2}+2\omega^{2}+\gamma^{2}_{g}\right)X^{2}\allowbreak+\allowbreak\allowbreak\omega^{4}
\end{equation}
which can be written, by defining $T=X^{2}$, as 
\begin{equation}
T^{2}+\left(-\gamma^{2}+2\omega^{2}+\gamma^{2}_{g}\right)T+\allowbreak\allowbreak\omega^{4} ,
\end{equation}
and has the eigenvalues 
\begin{equation}
\left(\begin{array}{c}
\overset{-}{}\lambda_{1}=\frac{1}{2}\gamma^{2}-\omega^{2}-\frac{1}{2}\gamma^{2}_{g}-\frac{1}{2}\sqrt{\left(\gamma^{2}-\gamma^{2}_{g}\right)\left(\gamma^{2}-4\omega^{2}-\gamma^{2}_{g}\right)}\\
\overset{-}{}\lambda_{2}=\frac{1}{2}\gamma^{2}-\omega^{2}-\frac{1}{2}\gamma^{2}_{g}+\frac{1}{2}\sqrt{\left(\gamma^{2}-\gamma^{2}_{g}\right)\left(\gamma^{2}-4\omega^{2}-\gamma^{2}_{g}\right)}
\end{array}\right)
\end{equation}

To simplify the notation, one defines 
\begin{equation}
a^{2}=\omega^{2}-\frac{\gamma^{2}}{4}+\frac{\gamma^{2}_{g}}{4},
\end{equation}
and the eigenvalues can be written as 
\begin{equation}
\left(\begin{array}{c}
\overset{-}{\lambda}_{1}=-\frac{1}{4}\left(2a-\sqrt{\gamma^{2}_{g}-\gamma^{2}}\right)^{2}\\
\overset{-}{\lambda}_{2}=-\frac{1}{4}\left(2a+\sqrt{\gamma^{2}_{g}-\gamma^{2}}\right)^{2}
\end{array}\right).
\end{equation}

Finally, the four roots of the characteristic polynomial of $A$ are
\begin{equation}
\left(\begin{array}{c}
\lambda_{1}=i\frac{1}{2}\left(2a-\sqrt{\gamma^{2}_{g}-\gamma^{2}}\right)\\
\lambda_{2}=i\frac{1}{2}\left(2a+\sqrt{\gamma^{2}_{g}-\gamma^{2}}\right)\\
\lambda_{3}=\lambda^{\ast}_{1}\\
\lambda_{4}=\lambda^{\ast}_{2}
\end{array}\right)
\end{equation}

The corresponding eigenvectors are certainly complex conjugate of each other, and
this allows one to restrict the space of states to the complex space
$C^{2}$. Consequently, we write the complexified equation in a
similar way that we presented for the case of only one circuit, restricting
the analysis to 
\begin{equation}
\frac{d}{dt}\left[\begin{array}{c}
z_{1}\\
z_{2}
\end{array}\right]=\left[\begin{array}{cc}
i\frac{1}{2}\left(2a+\sqrt{\gamma^{2}_{g}-\gamma^{2}}\right) & 0\\
0 & i\frac{1}{2}\left(2a-\sqrt{\gamma^{2}_{g}-\gamma^{2}}\right)
\end{array}\right]\left[\begin{array}{c}
z_{1}\\
z_{2}
\end{array}\right] ,
\end{equation}

which, as before, can be seen as an equivalent Schrödinger equation
with the Hamiltonian 
\begin{equation}
H=\left[\begin{array}{cc}
-\frac{1}{2}\left(2a+\sqrt{\gamma^{2}_{g}-\gamma^{2}}\right) & 0\\
0 & -\frac{1}{2}\left(2a-\sqrt{\gamma^{2}_{g}-\gamma^{2}}\right)
\end{array}\right] .
\end{equation}

This expression has successfully completed the complexification of
the initial classical case. Just to make more clear the connection
between the coupled classical circuits and the $PT$-symmetric conventional
expression of the Hamiltonian, we next perform  a similarity transformation.

To do so, let us start recalling the characteristic polynomial of
the matrix $H$ 
\begin{equation}
X^{2}+2aX+a^{2}+\frac{\gamma^{2}}{4}-\frac{\gamma^{2}_{g}}{4}.
\end{equation}
Then, we can propose a general $2\times2$ matrix $M$, so that it
reproduces the same characteristic polynomial as above 
\begin{equation}
M=\left[\begin{array}{cc}
x_{1} & x_{3}\\
x_{4} & x_{2}
\end{array}\right],
\end{equation}
and try to determine its parameters by taking into account that, for this
general matrix, the characteristic polynomial is 
\begin{equation}
X^{2}+\left(-x_{1}-x_{2}\right)X+\left(x_{1}x_{2}-x_{3}x_{4}\right)\label{cp} ,
\end{equation}
which allows one to identify its  coefficients with those of Eq. (\ref{cp}) 
\begin{equation}
\left(x_{1}+x_{2}\right)=-2a .
\end{equation}
One can choose 
\begin{equation}
x_{1}=-a-i\frac{\gamma}{4}\,\,\,\,;\,\,\,\,x_{2}=-a+i\frac{\gamma}{4} .
\end{equation}

On the other hand, from the independent term of the polynomial 
\begin{equation}
a^{2}+\frac{\gamma^{2}}{4}-\frac{\gamma^{2}_{g}}{4}=\left(x_{1}x_{2}-x_{3}x_{4}\right)=a^{2}+\gamma^{2}-x_{3}x_{4}
\end{equation}
we get
\begin{equation}
\gamma^{2}_{g}=x_{3}x_{4} ,
\end{equation}
which is verified if 
\begin{equation}
x_{3}=i\,\gamma_{g}\,\,\,\,;\,\,\,x_{4}=-i\,\gamma_{g} .
\end{equation}

Finally, the similarity transformed Hamiltonian results in
\begin{equation}
H=\left[\begin{array}{cc}
-a-i\gamma & i\gamma_{g}\\
-i\gamma_{g} & -a+i\gamma
\end{array}\right] ,
\end{equation}
which clearly displays the usual form of a $PT$-symmetric Hamiltonian

From the above examples, it is clear how to generalize the complexification
procedure to a general finite dimension. Certainly, the required algebraic
manipulations could be quite cumbersome, so that a more general
approach is needed for a larger electric circuit. Such approach is
addressed in the following section.

\section{Quantum simulation of differential equations}\label{SCH}

The examples discussed in the previous section can be brought to the
more general arena of simulating a system of differential equations
using quantum algorithms on quantum hardware. There have been several
proposals, with different complexity reduction, as compared to classical
methods. For a recent review, we refer the reader to \cite{raisuddin2024reviewquantumscientificcomputing}.
A specific quantum algorithm has been introduced to simulate the dynamics
of electrical circuits consisting of resistors, inductors, and capacitors
along with power sources, which provides an exponential speed-up over
classical algorithms \cite{dutt2026simulatingdynamicsrlccircuits}.

The complexification method offers a way to undergo this task, by
mapping the original real system into a Schrödinger equation. However,
the procedure quickly becomes cumbersome and impractical as the size
of the system is increased, and one would like to design a more straightforward
path that allows for such direct mapping. Here, we wish to emphasize
a method that aligns well with the ideas developed in previous sections,
in the sense that it aims for the most direct possible conversion
of a system of differential equations into an evolution governed by
the Schrödinger equation, using a Hamiltonian that includes an additional
variable. Indeed, the authors describe the method as a process of
\textquotedbl Schrödingerization'' \cite{PhysRevLett.133.230602}.
This method ends with equations that can be handled even in involved
cases by numerical simulations, an has been applied to finding the
numerical solution in some practical cases, such as the heat, convection,
Fokker-Planck, linear Boltzmann, and Black-Scholes equations \cite{PhysRevA.108.032603}.
Here we relate this procedure to the problem of simulating an electric
circuit, which provides an additional test for the Schrödingerization
method.

%The comparison of the results of the latter method with those obtained by means of the previously presented 
%procedure provides a cross check for both proposals.

We briefly summarize
the Schrödingerization procedure. Let us consider the following equation
\begin{equation}
\frac{\partial}{\partial t}\ket{u(t)}=A\ket{u(t)},\label{eq:ODE}
\end{equation}
where $\ket{u(t)}$ is a vector on an ``internal'' $N$- dimensional
Hilbert space $\mathcal{H}_{I}$, and $A$ is a $N\times N$ matrix.
The Schrödingerization first decomposes it into the Hermitian and
anti-Hermitian parts: 
\begin{equation}
A=H_{1}+iH_{2},\qquad H_{1}=(A+A^{\dagger})/2,\qquad H_{2}=(A-A^{\dagger})/(2i),\label{eq:Adecomposition}
\end{equation}
It next defines a warp transformation 
\begin{equation}
\ket{\omega(p,t)}=e^{-p}\ket{u(t)}.\label{eq:warp}
\end{equation}
The new variable $p$ has to be interpreted as a new degree of freedom
associated to an extra Hilbert space $\mathcal{H}_{p}$, so that $\ket{\omega(p,t)}$
belongs to the composed space $H=\mathcal{H}_{I}\otimes\mathcal{H}_{p}$.
For the transformed vector one readily obtains 
\begin{equation}
i\frac{\partial}{\partial t}\ket{\omega(p,t)}=H_{1}(-i\frac{\partial}{\partial p})\ket{\omega(p,t)}-H_{2}\ket{\omega(p,t)}.\label{eq:Schrodpt}
\end{equation}
Notice that, in the above equation, the derivative in $p$ can be
interpreted as a momentum operator $P_p=-i\frac{\partial}{\partial p}$
acting on the $\mathcal{H}_{p}$ space. The main point is that Eq.
(\ref{eq:Schrodpt}) describes an unitary evolution, even if the initial
$(\ref{eq:ODE})$ was not. In the following subsections we discuss
how Eq. (\ref{eq:Schrodpt}) can be solved on a quantum computer for the
case of just one simple oscillator, although it can be easily generalized
to more complicated circuits.

\subsection{Quantum circuit implementation}

For an implementation on a quantum computer, the auxiliary variable
$p$ has to be represented by a finite-dimensional quantum register.
We truncate the auxiliary domain to the finite interval

\begin{equation}
p\in[-L/2,L/2),
\end{equation}

and discretize it using

\begin{equation}
N_{p}=2^{n_{p}}
\end{equation}

grid points. The grid points are defined as

\begin{equation}
p_{j}=-\frac{L}{2}+j\Delta p,\qquad j=0,\ldots,N_{p}-1,
\end{equation}

with

\begin{equation}
\Delta p=\frac{L}{N_{p}}.
\end{equation}

The auxiliary Hilbert space is then spanned by the computational basis
states

\begin{equation}
\ket j_{p},\qquad j=0,\ldots,N_{p}-1,
\end{equation}

which encode the discretized values $p_{j}$.

An internal space of dimension
$N$ requires $\lceil\log_{2}N\rceil$ qubits, while the auxiliary
register requires $n_{p}=\log_{2}N_{p}$ qubits.

In the discretized representation, the enlarged state is written as

\begin{equation}
\ket{\omega(t)}=\sum^{N-1}_{\alpha=0}\sum^{N_{p}-1}_{j=0}\omega_{\alpha,j}(t)\ket{\alpha}\ket j_{p}.
\end{equation}

Here, $\ket{\alpha}$ labels the internal degrees of freedom, while
$\ket j_{p}$ labels the auxiliary coordinate.

On the finite grid, $P_p$ is diagonal in the discrete Fourier basis,
with momentum values

\begin{equation}
k_{n}=\frac{2\pi}{L}\nu_{n},\qquad\nu_{n}=0,1,\ldots,\frac{N_{p}}{2}-1,-\frac{N_{p}}{2},\ldots,-1,
\end{equation}

for even $N_{p}$. Defining

\begin{equation}
D_{k}=\operatorname{diag}(k_{n}),
\end{equation}

the momentum operator can be written, using the QFT convention adopted
in \texttt{PennyLane}~\cite{bergholm2022pennylaneautomaticdifferentiationhybrid},
as

\begin{equation}
P_{p}=\mathrm{QFT}_{N_{p}}D_{k}\mathrm{QFT}^{\dagger}_{N_{p}},
\end{equation}
or equivalently, 
\begin{equation}
\mathrm{QFT}^{\dagger}_{N_{p}}P_{p}\mathrm{QFT}_{N_{p}}=D_{k}.
\end{equation}

Therefore, $\mathrm{QFT}^{\dagger}_{N_{p}}$ transforms the auxiliary
register from the $p$ basis to the momentum basis, where the momentum
operator is diagonal. After the evolution, $\mathrm{QFT}_{N_{p}}$
transforms the register back to the $p$ basis.

Using this representation, the Schrödingerized Hamiltonian

\begin{equation}
H_{\mathrm{Sch}}=H_{1}\otimes P_{p}-H_{2}\otimes I_{p}
\end{equation}

can be implemented through the equivalent momentum-basis Hamiltonian

\begin{equation}
H_{\mathrm{mom}}=H_{1}\otimes D_{k}-H_{2}\otimes I_{p}.\label{eq:Hmom}
\end{equation}

The complete time evolution is then given by

\begin{equation}
\ket{\omega(t)}=\left(I\otimes\mathrm{QFT}_{N_{p}}\right)e^{-iH_{\mathrm{mom}}t}\left(I\otimes\mathrm{QFT}^{\dagger}_{N_{p}}\right)\ket{\omega(0)}.\label{eq:qft_evolution}
\end{equation}

Thus, the Fourier transformation is performed coherently on the auxiliary
quantum register through the QFT rather than as a classical preprocessing
step.

\paragraph{Single-oscillator implementation.}

We now apply the Schrödingerization procedure to the single damped
oscillator introduced in the previous section, see Eq. (\ref{eq:oneoscillator}). 

For the numerical example considered here, we take

\begin{equation}
\omega=2,\qquad\gamma=1,
\end{equation}

so that

\begin{equation}
A=\left[\begin{array}{cc}
0 & 1\\
-4 & -2
\end{array}\right].
\end{equation}

The eigenvalues of $A$ are $-1\pm i\sqrt{3},$ corresponding to a
stable underdamped oscillator.

The exact classical solution used for comparison is

\begin{equation}
u_{\mathrm{exact}}(t)=e^{At}u_{0}.\label{eq:exact_oscillator}
\end{equation}

With initial condition

\begin{equation}
u_{0}=\left[\begin{array}{c}
1\\
0
\end{array}\right],
\end{equation}
one readily obtains $u_{\mathrm{exact}}(t)=e^{-t}(\frac{\sin\sqrt{3}t}{\sqrt{3}}+\cos\sqrt{3}t,-\frac{4\sin\sqrt{3}t}{\sqrt{3}})^{T}$.

The decomposition defined in Eq. (\ref{eq:Adecomposition}) gives 

\begin{equation}
H_{1}=\left[\begin{array}{cc}
0 & -\frac{3}{2}\\
-\frac{3}{2} & -2
\end{array}\right],
\end{equation}

and

\begin{equation}
H_{2}=\left[\begin{array}{cc}
0 & -\frac{5}{2}i\\
\frac{5}{2}i & 0
\end{array}\right].
\end{equation}

Since the internal space has dimension two, a single qubit is sufficient
to encode the oscillator variables.

The lifted initial condition is chosen as

\begin{equation}
\ket{\omega(0)}=\ket{u_{0}}\otimes\ket g,
\end{equation}

where the auxiliary state $\ket g$ encodes the symmetrized profile

\begin{equation}
g(p_{j})=e^{-|p_{j}|}.
\end{equation}

Since the state must be normalized for quantum state preparation,
we define

\begin{equation}
\ket g=\frac{1}{\mathcal{N}_{g}}\sum^{N_{p}-1}_{j=0}e^{-|p_{j}|}\ket j_{p},
\end{equation}

where

\begin{equation}
\mathcal{N}_{g}=\left(\sum^{N_{p}-1}_{j=0}e^{-2|p_{j}|}\right)^{1/2}.
\end{equation}

\paragraph{Choice of the auxiliary domain and recovery.}

The auxiliary domain must be sufficiently large for two reasons. First,
the recovery point must lie inside the discretized interval. Second,
the artificial periodic boundary introduced by the Fourier representation
must be sufficiently far from the relevant region so that the auxiliary
profile has decayed to a negligible value.

Following the Schrödingerization prescription, the original solution
can be recovered from

\begin{equation}
\ket{u(t)}=e^{p}\ket{\omega(p,t)}
\end{equation}

provided that $p$ is chosen within the appropriate recovery region.
Notice that, in this example, the eigenvalues of $H_{1}$ are $\lambda_{1}=-(1+\frac{\sqrt{13}}{2})\simeq-2.80278$,
and $\lambda_{2}=\frac{1}{2}\left(\sqrt{13}-2\right)\simeq0.802776$.
As discussed in \cite{Jin:2024qsi}, the existence of non-negative
eigenvalues of $H_{1}$ requires special attention for the recovery
process to work. In our implementation, a fixed grid point satisfying

\begin{equation}
p_{j}\geq p^{\diamond}
\end{equation}

is used for all evolution times, where

\begin{equation}
p^{\diamond}=\lambda^{+}_{\max}(H_{1})T.\label{eq:pdiamond}
\end{equation}
Here, $T$ denotes the final evolution time and $\lambda^{+}_{\max}(H_{1})$
is the largest positive eigenvalue of $H_{1}$. For the oscillator
considered here, 
\begin{equation}
\lambda^{+}_{\max}(H_{1})=\lambda_{2}.
\end{equation}
For a total evolution time 
\begin{equation}
T=6,
\end{equation}
the corresponding recovery threshold is 
\begin{equation}
p^{\diamond}\simeq4.8167.
\end{equation}

The length of the auxiliary interval is chosen such that the exponentially
decaying auxiliary profile is negligible at the boundaries. In the
numerical implementation, $L$ is determined from the condition 
\begin{equation}
e^{-L/2+\lambda_{\max}(H_{1})T}\simeq\varepsilon,\label{eq:L_condition}
\end{equation}
where $\varepsilon$ is a prescribed numerical tolerance. We use 
\begin{equation}
\varepsilon=10^{-6}.
\end{equation}
This choice keeps the recovery region well inside the discretized
interval while suppressing the influence of the artificial periodic
boundary.

After the time evolution, the amplitudes of the Schrödingerized state
are reshaped into the matrix

\begin{equation}
W(t)=\left[\omega_{\alpha,j}(t)\right]\in\mathbb{C}^{2\times N_{p}},
\end{equation}

where the first index labels the internal oscillator variables and
the second index labels the auxiliary grid points. The recovered solution
is then obtained from the column associated with a grid point in the
recovery region,

\begin{equation}
u_{\mathrm{rec}}(t)=\frac{W_{:,j}(t)}{g(p_{j})},\qquad p_{j}\geq p^{\diamond}.\label{eq:recovery_numerical}
\end{equation}

For positive $p_{j}$, this procedure corresponds to undoing the exponential
warp factor used in the construction of the Schrödingerized state,
while also accounting for the normalization introduced in the preparation
of the auxiliary state.

\paragraph{Circuit implementation.}

The circuit implementation follows directly from Eq.~(\ref{eq:qft_evolution}).
In \texttt{PennyLane}, the corresponding sequence is

\begin{equation}
\mathrm{StatePrep}\quad\longrightarrow\quad\mathrm{QFT}^{\dagger}\quad\longrightarrow\quad e^{-iH_{\mathrm{mom}}t}\quad\longrightarrow\quad\mathrm{QFT}.
\end{equation}

\begin{figure}[t]
    \centering
    \includegraphics[width=0.6\textwidth]{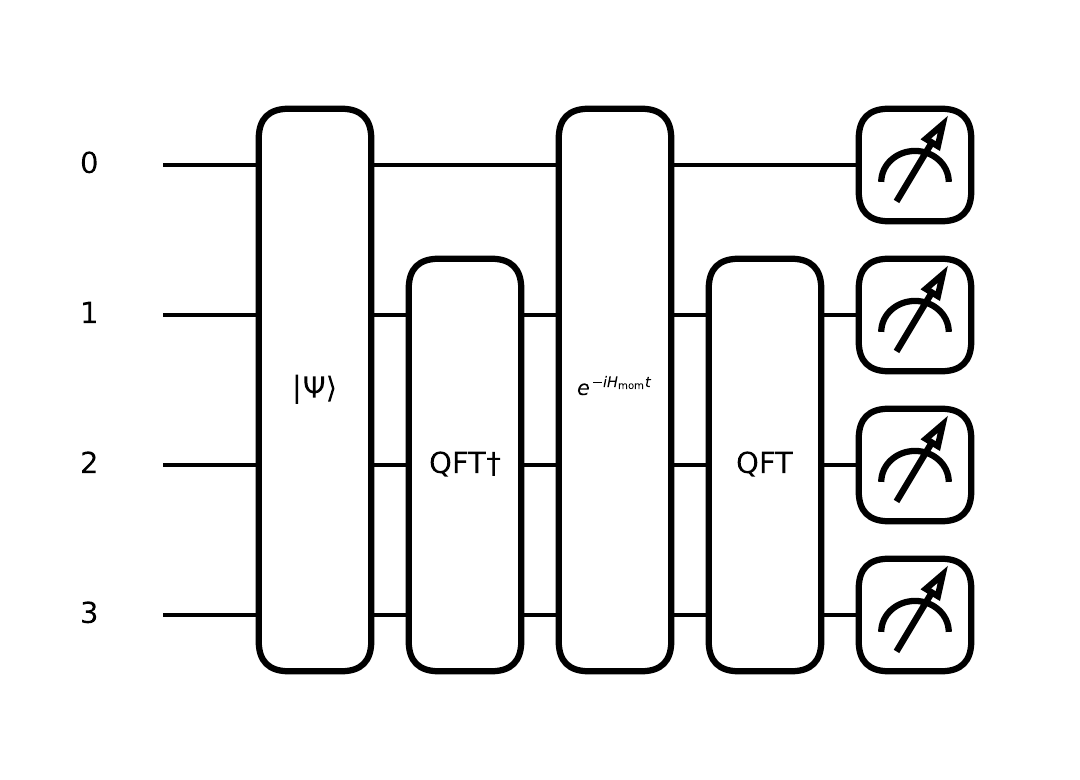}
    \caption{
    Schematic quantum circuit for the Schr\"odingerization of the single
    damped oscillator using $N_p=8$. The first qubit encodes the
    two-dimensional oscillator state, while the remaining three qubits encode
    the auxiliary variable $p$. The initial state
    $\ket{u_0}\otimes\ket{g}$ is prepared coherently, followed by an inverse
    QFT on the auxiliary register, evolution under the momentum-basis
    Hamiltonian $H_{\mathrm{mom}}$, and a final QFT that transforms the
    auxiliary register back to the $p$ basis. The numerical simulations use
    $N_p=2048$, corresponding to 11 auxiliary qubits and 12 qubits in total;
    the circuit shown here illustrates the analogous structure for a smaller
    auxiliary register.
    }
    \label{fig:schrodingerization_circuit}
\end{figure}

For the numerical simulations presented below, we use $N_{p}=2048$,
corresponding to an auxiliary register of $n_{p}=11$ qubits. Together
with the single qubit encoding the oscillator, the complete state
is represented using 12 qubits. Displaying the corresponding full
circuit would be unnecessarily cumbersome. Therefore, Fig.~\ref{fig:schrodingerization_circuit}  illustrates
the same construction for $N_{p}=8$, which requires three auxiliary
qubits and gives a four-qubit circuit in total. Increasing $N_{p}$
changes the size of the auxiliary register but not the structure of
the algorithm.

%%%%%%% Figure here for visual convenience

%%%%%%%

\paragraph{Implementation summary.}

The complete quantum implementation can therefore be summarized as
follows: 
\begin{enumerate}
\item The original differential equation is written in first-order form,
\begin{equation}
\frac{d}{dt}u(t)=Au(t),
\end{equation}
and its generator is decomposed as $A=H_{1}+iH_{2}$.
\item The auxiliary coordinate $p$ is discretized using $N_{p}=2^{n_{p}}$
grid points and encoded in an $n_{p}$-qubit register.
\item The initial state $\ket{u_{0}}\otimes\ket g$ is prepared, where $\ket g$
encodes the normalized auxiliary profile $e^{-|p|}$.
\item An inverse QFT is applied to the auxiliary register to transform the
state from the $p$ basis to the momentum basis.
\item The enlarged state evolved under the Hermitian momentum-basis Hamiltonian
\begin{equation}
H_{\mathrm{mom}}=H_{1}\otimes D_{k}-H_{2}\otimes I_{p}.
\end{equation}
\item A QFT is applied to transform the auxiliary register back to the $p$
basis.
\item The original solution is reconstructed from a grid point satisfying
$p_{j}\geq p^{\diamond}$ using Eq.~(\ref{eq:recovery_numerical}). 
\end{enumerate}

\subsection{Numerical results}

To assess the accuracy of the quantum implementation, the reconstructed
solution is compared with the exact classical evolution. We consider
separately the two components $x(t)$ and $y(t)=\dot{x}(t)$ and,
in addition, evaluate the normalized direction error 
\begin{equation}
\epsilon_{\mathrm{dir}}(t)=\left\Vert \frac{u_{\mathrm{exact}}(t)}{\left\Vert u_{\mathrm{exact}}(t)\right\Vert }-\frac{u_{\mathrm{rec}}(t)}{\left\Vert u_{\mathrm{rec}}(t)\right\Vert }\right\Vert _{2}.
\end{equation}

\begin{figure}[t]
\centering \includegraphics[width=1\textwidth]{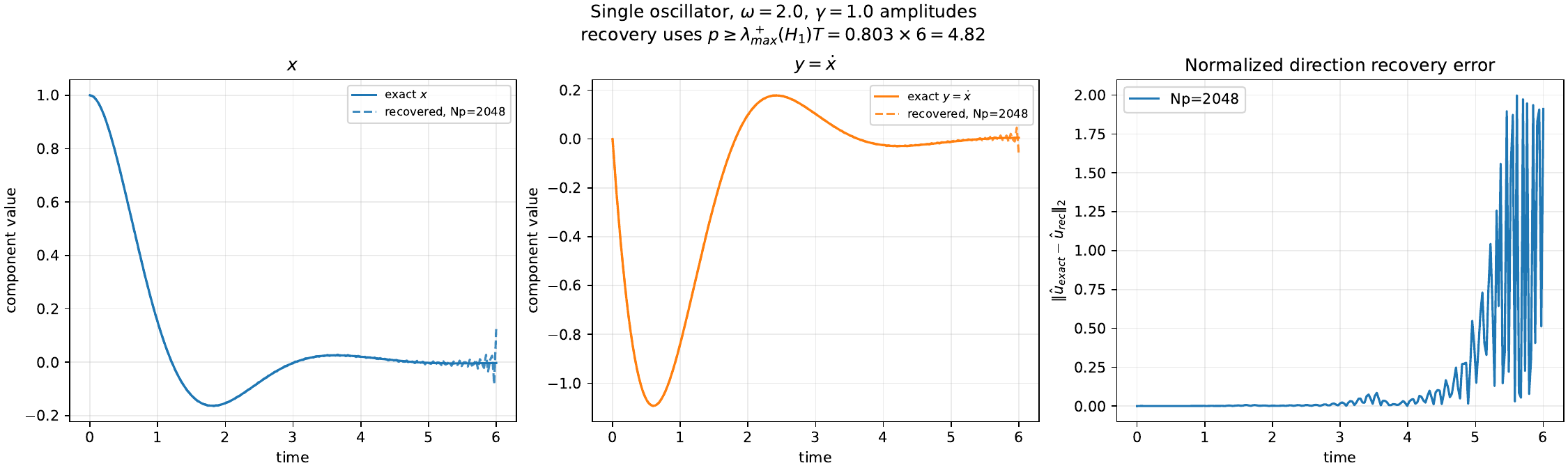}
\caption{ Numerical simulation of the damped harmonic oscillator using the
Schrödingerization procedure for $\omega=2$, $\gamma=1$, and $u_{0}=(1,0)^{T}$,
with $N_{p}=2048$. The left and central panels compare the reconstructed
Schrödingerized dynamics with the exact classical solution for $x(t)$
and $y(t)=\dot{x}(t)$, respectively, showing very good agreement
over most of the simulated time interval. The right panel shows the
normalized direction error, which remains small during the initial
and intermediate evolution and increases at late times, when the oscillator
amplitudes have strongly decayed and the recovery becomes more sensitive
to numerical errors. }
\label{fig:single_oscillator_results} 
\end{figure}

Figure~\ref{fig:single_oscillator_results} shows the numerical results
obtained for the single damped oscillator using $N_{p}=2048$. The
left and central panels compare the exact classical dynamics with
the solution recovered from the Schrödingerized evolution for $x(t)$
and $y(t)$, respectively. In both cases, the reconstructed solution
closely follows the exact evolution, with the two curves being practically
indistinguishable over most of the simulated time interval. In particular,
the method accurately reproduces both the oscillatory behavior and
the exponential damping of the original system.

Small deviations become visible only at late evolution times, when
the amplitudes of both components have already decayed to values close
to zero. The same behavior is reflected in the normalized direction
error shown in the right panel. The error remains small throughout
the initial and intermediate stages of the evolution and grows appreciably
only towards the end of the simulation.

The increase of the error at late times can be understood from the
recovery procedure. Since $H_{1}$ has a positive eigenvalue, the
solution must be reconstructed at a positive value $p_{j}\geq p^{\diamond}$.
 Recovery involves division by the exponentially decaying auxiliary
profile $g(p_{j})$, or equivalently multiplication by an exponentially
increasing factor. Consequently, small discretization and Fourier
errors in the Schrödingerized state can be amplified during the reconstruction.
This effect becomes particularly relevant at late times, when the
physical amplitudes of the damped oscillator are already very small.

Despite this increased numerical sensitivity in the late-time regime,
the results show very good agreement between the reconstructed and
exact solutions over most of the evolution, demonstrating that the
QFT-based Schrödingerization implementation successfully reproduces
the classical dynamics of the damped oscillator.

\section{Conclusions}

Rephrasing classical differential equations in a way that can be directly approached using a quantum hardware is a subject of paramount interest, as it can be applied to many classical systems in an efficient way. In recent years, there has been a large effort in this direction. For a recent review, we refer the reader to \cite{raisuddin2024reviewquantumscientificcomputing}. 

In this paper, we first addressed the above goal as a general complexification procedure, that connects a classical dynamics with an equivalent quantum one. 
To illustrate this procedure, we discuss two examples of classical differential equations related to electric circuits that can be exactly solved. 
However, the complexification method becomes rapidly untractable as the size of the circuit increases, so one has to find an alternative. Among the ones suggested in the literature, we find very appealing the Schr\"odingerization procedure \cite{PhysRevLett.133.230602,PhysRevA.108.032603}, as it straightforwardly maps the initial classical system of equations into a quantum system which is governed by a Schrödinger equation, by adding an extra degree of freedom, which can be treated as a momentum extra space via a QFT. Given this important connection, we have explored its performance for the case of a simple oscillator, provided that some care is taken for the treatment of the momentum variable. When this done, we find a good agreement with the exact solution. We think that QFT-based Schr\"odingerization procedure appears to be a promising way for quantum computing tools to be used in solving classical differential equations.

\begin{acknowledgments}
A.P. received support through Project No. PID2023-152724NA-I00, with
funding from MCIU/AEI/10.13039/501100011033 and FSE+, Generalitat
Valenciana Grant No. CIPROM/2022/66, the Severo Ochoa Grant No. CEX2023-001292-S,
the Ministry of Economic Affairs and Digital Transformation of the
Spanish Government through the QUANTUM ENIA project QUANTUM SPAIN,
and the European Union through the Recovery, Transformation and Resilience
Plan, NextGenerationEU, within the framework of the Digital Spain
2026 Agenda, and the CSIC Interdisciplinary Thematic Platform on Quantum
Technologies. This project also received funding from the European
Union’s Horizon 2020 research and innovation program under Grant Agreement
CaLIGOLA No. MSCA-2021-SE-01-101086123. CGC warmly acknowledges the
hospitality of the Departamento de Física Teórica, Universidad de
Valencia, under Grant Agreement CaLIGOLA No. MSCA-2021-SE-01-101086123,
where part of this work was discussed. RGL is funded by the grant CIACIF/2021/136
from Generalitat Valenciana. 
\end{acknowledgments}

\bibliography{Bibliography}

@Article{doi:10.1098/rspa.2016.0615,
  author   = {Caruso, M. and Fanchiotti, H. and García Canal, C. A. and Mayosky, M. and Veiga, A.},
  journal  = {Proceedings of the Royal Society A: Mathematical, Physical and Engineering Sciences},
  title    = {The quantum CP-violating kaon system reproduced in the electronic laboratory},
  year     = {2016},
  number   = {2195},
  pages    = {20160615},
  volume   = {472},
  doi      = {10.1098/rspa.2016.0615},
  eprint   = {https://royalsocietypublishing.org/doi/pdf/10.1098/rspa.2016.0615},
  url      = {https://royalsocietypublishing.org/doi/abs/10.1098/rspa.2016.0615},
}

@book{Arnold2012,
  author    = {Arnold, V. I.},
  editor    = {Levi, Mark},
  title     = {Geometrical Methods in the Theory
               of Ordinary Differential Equations},
  publisher = {Springer New York},
  year      = {2012},
  edition   = {2},
  series    = {Grundlehren der mathematischen Wissenschaften},
  volume    = {250},
  isbn      = {9781461210375},
  doi       = {10.1007/978-1-4612-1037-5},
  note      = {Translated by J. Szücs}
}

@book{bookmath, author = {Treil, Sergei}, title = {Linear Algebra Done Wrong}, publisher = {Flooved.com on behalf of the author}, year = {2011}, url = {https://archive.org/details/flooved3458} }

@Article{Fanchiotti2023,
  author   = {Fanchiotti, H. and García Canal, C. A. and Mayosky, M. and Veiga, A. and Vento, V.},
  journal  = {Brazilian Journal of Physics},
  title    = {The Geometric Phase in Classical Systems and in the Equivalent Quantum Hermitian and Non-Hermitian PT-Symmetric Systems},
  year     = {2023},
  issn     = {1678-4448},
  number   = {6},
  pages    = {143},
  volume   = {53},
  doi      = {10.1007/s13538-023-01350-5},
  refid    = {Fanchiotti2023},
  url      = {https://doi.org/10.1007/s13538-023-01350-5},
}

@Article{Fanchiotti2022,
  author   = {Fanchiotti, H. and García Canal, C. A. and Mayosky, M. and Veiga, A. and Vento, V.},
  journal  = {American Journal of Physics},
  title    = {Measuring the Hannay geometric phase},
  year     = {2022},
  issn     = {0002-9505},
  month    = {06},
  number   = {6},
  pages    = {430-435},
  volume   = {90},
  doi      = {10.1119/5.0081149},
  eprint   = {https://pubs.aip.org/aapt/ajp/article-pdf/90/6/430/20100684/430\_1\_5.0081149.pdf},
  url      = {https://doi.org/10.1119/5.0081149},
}

@book{arnold1992ode,
  title={Ordinary Differential Equations},
  author={Arnold, Vladimir I.},
  year={1992},
  publisher={Springer-Verlag},
  address={Berlin, Heidelberg},
  edition={3rd},
  isbn={978-3-540-34563-3},
  url={https://link.springer.com/book/9783540345633}
}

@article{PhysRevLett.133.230602,
  title = {Quantum Simulation of Partial Differential Equations via Schr\"odingerization},
  author = {Jin, Shi and Liu, Nana and Yu, Yue},
  journal = {Phys. Rev. Lett.},
  volume = {133},
  issue = {23},
  pages = {230602},
  numpages = {6},
  year = {2024},
  month = {Dec},
  publisher = {American Physical Society},
  doi = {10.1103/PhysRevLett.133.230602},
  url = {https://link.aps.org/doi/10.1103/PhysRevLett.133.230602}
}

@article{Telle48,
    author = {Bernard DH. Tellegen},
    title = {The gyrator, a new electric network element},
    journal = {Philips Res. Rep},
    volume={3},
    issue={2},
    pages={81-101},
    year ={1948}
}

@article{Jin:2024qsi,
    author = "Jin, Shi and Liu, Nana and Ma, Chuwen",
    title = {{On Schr{\"o}dingerization-Based Quantum Algorithms for Linear Dynamical Systems with Inhomogeneous Terms}},
    eprint = "2402.14696",
    archivePrefix = "arXiv",
    primaryClass = "math.NA",
    doi = "10.1137/24m164272x",
    journal = "SIAM J. Numer. Anal.",
    volume = "63",
    number = "4",
    pages = "1861--1885",
    year = "2025"
}

@Misc{miessen2025digitalquantumsimulationmanybody,
  author        = {Alexander Miessen},
  title         = {Digital quantum simulation of many-body systems: Making the most of intermediate-scale, noisy quantum computers},
  year          = {2025},
  archiveprefix = {arXiv},
  eprint        = {2508.21504},
  primaryclass  = {quant-ph},
  url           = {https://arxiv.org/abs/2508.21504},
}

@Article{https://doi.org/10.1002/wcms.70020,
  author   = {Zhang, Yukun and Zhang, Xiaoming and Sun, Jinzhao and Lin, Heng and Huang, Yifei and Lv, Dingshun and Yuan, Xiao},
  journal  = {WIREs Computational Molecular Science},
  title    = {Quantum Algorithms for Quantum Molecular Systems: A Survey},
  year     = {2025},
  note     = {e70020 CMS-1125.R1},
  number   = {3},
  pages    = {e70020},
  volume   = {15},
  doi      = {https://doi.org/10.1002/wcms.70020},
  eprint   = {https://wires.onlinelibrary.wiley.com/doi/pdf/10.1002/wcms.70020},
  url      = {https://wires.onlinelibrary.wiley.com/doi/abs/10.1002/wcms.70020},
}

@Article{Cirac2012,
  author   = {Cirac, J. Ignacio and Zoller, Peter},
  journal  = {Nature Physics},
  title    = {Goals and opportunities in quantum simulation},
  year     = {2012},
  issn     = {1745-2481},
  number   = {4},
  pages    = {264--266},
  volume   = {8},
  doi      = {10.1038/nphys2275},
  refid    = {Cirac2012},
  url      = {https://doi.org/10.1038/nphys2275},
}

@Article{Tennie2025,
  author   = {Tennie, Felix and Laizet, Sylvain and Lloyd, Seth and Magri, Luca},
  journal  = {Nature Reviews Physics},
  title    = {Quantum computing for nonlinear differential equations and turbulence},
  year     = {2025},
  issn     = {2522-5820},
  number   = {4},
  pages    = {220--230},
  volume   = {7},
  doi      = {10.1038/s42254-024-00799-w},
  refid    = {Tennie2025},
  url      = {https://doi.org/10.1038/s42254-024-00799-w},
}

@Article{PhysRevResearch.2.043102,
  author    = {Joseph, Ilon},
  journal   = {Phys. Rev. Res.},
  title     = {Koopman--von Neumann approach to quantum simulation of nonlinear classical dynamics},
  year      = {2020},
  month     = {Oct},
  pages     = {043102},
  volume    = {2},
  doi       = {10.1103/PhysRevResearch.2.043102},
  issue     = {4},
  numpages  = {17},
  publisher = {American Physical Society},
  url       = {https://link.aps.org/doi/10.1103/PhysRevResearch.2.043102},
}

@Book{Manin1980,
  author    = {Manin, Yuri I.},
  publisher = {Sov.Radio},
  title     = {Vychislimoe i nevychislimoe},
  year      = {1980},
  address   = {Moscow},
  note      = {},
  language  = {Russian},
  pages     = {13--15},
  title_en  = {Computable and Noncomputable},
  url       = {https://web.archive.org/web/20130510173823/http://publ.lib.ru/ARCHIVES/M/MANIN_Yuriy_Ivanovich/Manin_Yu.I._Vychislimoe_i_nevychislimoe.(1980).},
}

@Article{Feynman1982,
  author  = {Feynman, Richard P.},
  journal = {International Journal of Theoretical Physics},
  title   = {Simulating physics with computers},
  year    = {1982},
  issn    = {1572-9575},
  number  = {6},
  pages   = {467--488},
  volume  = {21},
  doi     = {10.1007/BF02650179},
  refid   = {Feynman1982},
  url     = {https://doi.org/10.1007/BF02650179},
}

@Misc{dutt2026simulatingdynamicsrlccircuits,
  author        = {Arkopal Dutt and Anirban Chowdhury and Kristan Temme and Hari Krovi},
  title         = {Simulating dynamics of RLC circuits with a quantum differential-algebraic equations solver},
  year          = {2026},
  archiveprefix = {arXiv},
  eprint        = {2604.26945},
  primaryclass  = {quant-ph},
  url           = {https://arxiv.org/abs/2604.26945},
}

@Misc{raisuddin2024reviewquantumscientificcomputing,
  author        = {Osama Muhammad Raisuddin and Suvranu De},
  title         = {A Review of Quantum Scientific Computing Algorithms for Engineering Problems},
  year          = {2024},
  archiveprefix = {arXiv},
  eprint        = {2408.13943},
  primaryclass  = {quant-ph},
  url           = {https://arxiv.org/abs/2408.13943},
}

@Article{PhysRevA.108.032603,
  author    = {Jin, Shi and Liu, Nana and Yu, Yue},
  journal   = {Phys. Rev. A},
  title     = {Quantum simulation of partial differential equations: Applications and detailed analysis},
  year      = {2023},
  month     = {Sep},
  pages     = {032603},
  volume    = {108},
  doi       = {10.1103/PhysRevA.108.032603},
  issue     = {3},
  numpages  = {20},
  publisher = {American Physical Society},
  url       = {https://link.aps.org/doi/10.1103/PhysRevA.108.032603},
}

@misc{bergholm2022pennylaneautomaticdifferentiationhybrid,
      title={PennyLane: Automatic differentiation of hybrid quantum-classical computations}, 
      author={Ville Bergholm and Josh Izaac and Maria Schuld and Christian Gogolin and Shahnawaz Ahmed and Vishnu Ajith and M. Sohaib Alam and Guillermo Alonso-Linaje and B. AkashNarayanan and Ali Asadi and Juan Miguel Arrazola and Utkarsh Azad and Sam Banning and Carsten Blank and Thomas R Bromley and Benjamin A. Cordier and Jack Ceroni and Alain Delgado and Olivia Di Matteo and Amintor Dusko and Tanya Garg and Diego Guala and Anthony Hayes and Ryan Hill and Aroosa Ijaz and Theodor Isacsson and David Ittah and Soran Jahangiri and Prateek Jain and Edward Jiang and Ankit Khandelwal and Korbinian Kottmann and Robert A. Lang and Christina Lee and Thomas Loke and Angus Lowe and Keri McKiernan and Johannes Jakob Meyer and J. A. Montañez-Barrera and Romain Moyard and Zeyue Niu and Lee James O'Riordan and Steven Oud and Ashish Panigrahi and Chae-Yeun Park and Daniel Polatajko and Nicolás Quesada and Chase Roberts and Nahum Sá and Isidor Schoch and Borun Shi and Shuli Shu and Sukin Sim and Arshpreet Singh and Ingrid Strandberg and Jay Soni and Antal Száva and Slimane Thabet and Rodrigo A. Vargas-Hernández and Trevor Vincent and Nicola Vitucci and Maurice Weber and David Wierichs and Roeland Wiersema and Moritz Willmann and Vincent Wong and Shaoming Zhang and Nathan Killoran},
      year={2022},
      eprint={1811.04968},
      archivePrefix={arXiv},
      primaryClass={quant-ph},
      url={https://arxiv.org/abs/1811.04968}, 
}

\end{document}